\documentclass [11pt,a4paper] {article}

\usepackage[cp1252]{inputenc}
\usepackage{amssymb}
\usepackage{amsmath}
\usepackage{amsfonts,amssymb}
\usepackage[dvips]{graphicx}
\usepackage{bbm}
\usepackage{enumerate}
\usepackage{amsthm}
\usepackage{cancel}

\DeclareMathAlphabet{\mathpzc}{OT1}{pzc}{m}{it}

\begin{document}

\title{An empiric logic approach to Einstein's version of the double-slit experiment}

\author{Arkady Bolotin\footnote{$Email: arkadyv@bgu.ac.il$\vspace{5pt}} \\ \textit{Ben-Gurion University of the Negev, Beersheba (Israel)}}

\maketitle

\begin{abstract}\noindent As per Einstein's design, particles are introduced into the double-slit experiment through a small hole in a plate which can either move up and down (and its momentum can be measured) or be stopped (and its position can be measured). Suppose one measures the position of the plate and this act verifies the statement that the interference pattern is observed in the experiment. However, if it is possible to think about the outcome that one would have obtained if one had measured plate’s momentum instead of its position, then it is possible to consider, together with the aforesaid statement, another statement that each particle passes through either slit of the double-slit screen. Hence, the proposition affirming the wave-like behavior and the proposition affirming the particle-like behavior might be true together, which would imply that Bohr’s complementarity principle is incorrect. The analysis of Einstein’s design and ways to refute it based on an approach that uses exclusively assignments of the truth values to experimental propositions is presented in this paper.\\

\end{abstract}

\section{Introduction}  

\noindent According to Bohr's complementarity principle, one can observe either the wave-like behavior or the particle-like behavior, but not both simultaneously. The emblematical example of complementarity of these behaviors is the double-slit experiment, which, in the words of Feynman, ``has in it the heart of quantum mechanics'' \cite{Feynman}.\\

\noindent Within a mathematical model allowing one to reason about the truth or falsehood of experimentally verifiable statements, the double-slit experiment can be described in this way. Let $P_{\text{wave-like}}$ denote the proposition affirming the wave-like behavior (i.e., the statement saying that the interference pattern is observed in the experiment) and $P_{\text{particle-like}}$ denote the proposition affirming the particle-like behavior (i.e., the statement saying that a particle passes through either slit). Then it follows that, consistent with Bohr's principle, $P_{\text{particle-like}}$ cannot be true when $P_{\text{wave-like}}$ is true, and vice versa. That is, these propositions cannot have the value of true simultaneously. In symbols,\smallskip

\begin{equation} \label{CP} 
   {\big[\mkern-4.3mu\big[ 
      P_{\text{wave-like}}
   \big]\mkern-4.3mu\big]}_v
   \neq
   {\big[\mkern-4.3mu\big[ 
      P_{\text{particle-like}}
   \big]\mkern-4.3mu\big]}_v
   \;\;\;\;   ,
\end{equation}
\smallskip

\noindent where the double-bracket notation is used to denote a valuation $v$, i.e., a mapping from the set of propositions $\{P\}$ to the set of the truth values, true and false, renamed to 1 and 0 respectively:\smallskip

\begin{equation}  
   v
   :
   \;
   \{P\}
   \to
   \{1,0\}
   \;\;\;\;   .
\end{equation}
\smallskip

\noindent In 1927, during the $5^{\text{th}}$ Solvay Conference, Einstein designed of a modified version of the double-slit experiment intended to demonstrate the violation of (\ref{CP}) and, thus, the inconsistency of quantum mechanics \cite{Jammer}.\\

\begin{figure}[ht!]
   \centering
   \includegraphics[scale=0.5]{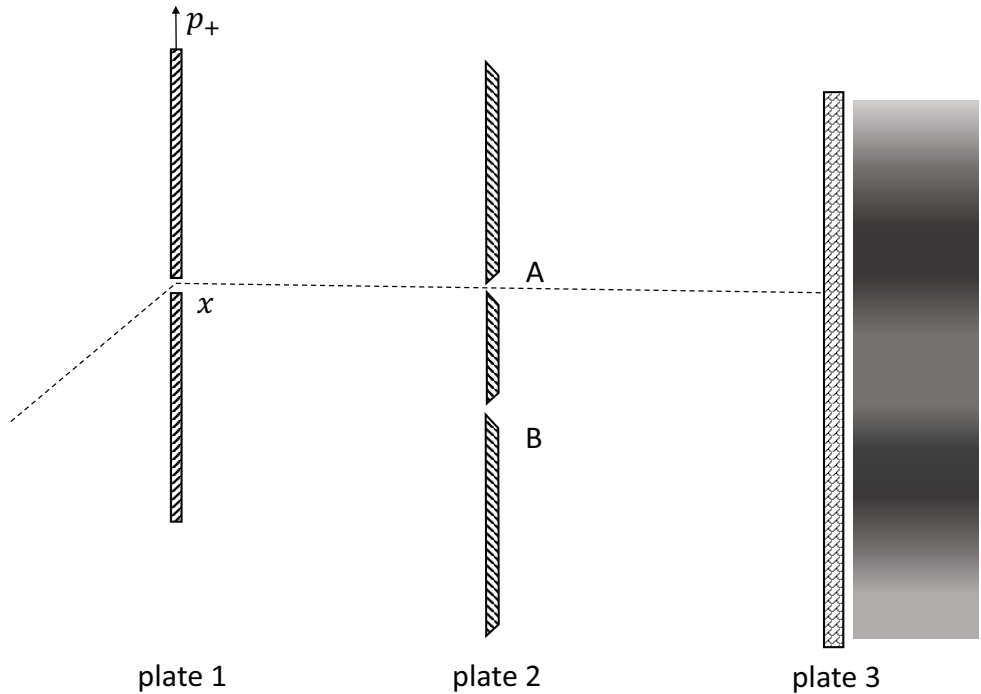}
   \caption{Einstein’s version of the double-slit experiment\label{fig1}}
\end{figure}

\noindent Imagine the double-slit experiment in which a plate that has a hole in it (depicted as plate 1 in Fig. \ref{fig1}) is added in front of the double-slit screen (plate 2). Plate 1 can move up and down and its momentum $p$ can be measured. If a particle coming through the hole in plate 1 is deflected in the direction of the double-slit screen, then the particle transfers a momentum to plate 1. By determining whether this momentum is positive or negative, one can make known the slit – either A or B – that the particle got through flying to plate 3. In contrast, if plate 1 is stopped and its position $x$ is measured, one can know exactly where the particle starts its route to plate 3.\\

\noindent Suppose that position $x$ of plate 1 is measured; in that case, the interference pattern can be observed, and so the proposition $P_{\text{wave-like}}$ can be verified. But if reliable measurements of counterfactual and definite kind are possible, then, together with the factual value of $x$, one may consider every imaginable value of momentum $p$ of plate 1. So, the proposition $P_{\text{particle-like}}$ can be thought as verified together with the proposition $P_{\text{wave-like}}$, that is,\smallskip

\begin{equation}  
   {\big[\mkern-4.3mu\big[ 
      P_{\text{wave-like}}
   \big]\mkern-4.3mu\big]}_v
   =
   {\big[\mkern-4.3mu\big[ 
      P_{\text{particle-like}}
   \big]\mkern-4.3mu\big]}_v
   =
   1
   \;\;\;\;   ,
\end{equation}
\smallskip

\noindent implying that the complementarity principle must be incorrect.\\

\noindent The purpose of this paper is to analyze Einstein’s design and ways to refute it using exclusively assignments of the truth values to experimental propositions.\\

\section{Definitions and preliminaries}  

\noindent In a mathematical model, which makes it possible for one to assign the value of true or false to experimentally verifiable (testable) statements relating to a quantum system, each proposition
$\!\!$\footnote{\label{f1}Even though the nature or existence of propositions as abstract meanings is still a matter of philosophical controversy, for the purposes of this paper, the phrases "statement" and "proposition" are used interchangeably.\vspace{5pt}} is uniquely represented by a closed linear subspace of a Hilbert space $\mathcal{H}$ associated with the system \cite{Mackey,Putnam}. Consequently, the proposition $P$ is assigned the value of true if and only if the system is in the pure state $|\Psi\rangle$ belonging to the subspace $\mathcal{H}_P$ that uniquely represents $P$ \cite{Redei}. Symbolically, this can be expressed as\smallskip

\begin{equation}  
   {\big[\mkern-4.3mu\big[ 
      P \left(
         |\Psi\rangle \in \mathcal{H}_P
      \right)
   \big]\mkern-4.3mu\big]}_v
   =
   1
   \;\;\;\;   ,
\end{equation}
\smallskip

\noindent where $P(|\Psi\rangle \in \mathcal{H}_P)$ stands for ``$P$ in the state $|\Psi\rangle \in \mathcal{H}_P$''. Otherwise, i.e., if the system is in the pure state that does not reside in $\mathcal{H}_P$, the proposition $P$ is assigned the value of false:\smallskip

\begin{equation}  
   {\big[\mkern-4.3mu\big[ 
      P \left(
         |\Psi\rangle \notin \mathcal{H}_P
      \right)
   \big]\mkern-4.3mu\big]}_v
   =
   0
   \;\;\;\;   .
\end{equation}
\smallskip

\noindent Together with the determined in such a way truth or falsehood of the proposition $P$, which can be labelled \textit{a factual truth-value} of $P$, one can consider \textit{a counterfactual truth-value} of $P$, namely, either of the truth values, 1 or 0, that would have been taken by $P$ if, instead of the state $|\Psi\rangle \in \mathcal{H}_P$ or $|\Psi\rangle \notin \mathcal{H}_P$, the system had been in the state $|\Psi\rangle \in \mathcal{H}_Q$ where $\mathcal{H}_Q$ is a closed linear subspace of the Hilbert space $\mathcal{H}$ that \textit{does not represent} $P$.\\

\noindent In order to have a particular meaning, the counterfactual truth-value of the proposition $P$ in the state $|\Psi\rangle \in \mathcal{H}_Q$, must depend on \textit{the lattice-theoretic ordering relation} $\le$ over the subspaces $\mathcal{H}_Q$ and $\mathcal{H}_P$.\\

\noindent For example, suppose that the subspaces $\mathcal{H}_Q$ and $\mathcal{H}_P$ are members of the same poset (partially ordered set) $\mathcal{L}$ and they are \textit{comparable} with each other, specifically, $\mathcal{H}_Q \le \mathcal{H}_P$. In this case, the state $|\Psi\rangle$ lying in the subspace $\mathcal{H}_Q$ belongs to the subspace $\mathcal{H}_P$ as well. Therefore, the counterfactual truth-value of the proposition $P$ in $|\Psi\rangle \in \mathcal{H}_Q$ can be treated the same as the factual truth-value of $P$ in $|\Psi\rangle \in \mathcal{H}_P$; explicitly,\smallskip

\begin{equation}  
   \{
      \mathcal{H}_Q
      ,
      \mathcal{H}_P
   \}
   \subseteq
   \mathcal{L}
   ,
   \>
   \mathcal{H}_Q
   \le
   \mathcal{H}_P
   \textnormal{:}
   \;\;
   {\big[\mkern-4.3mu\big[ 
      P \left(
         |\Psi\rangle \in \mathcal{H}_Q
      \right)
   \big]\mkern-4.3mu\big]}_v
    =
   {\big[\mkern-4.3mu\big[ 
      P \left(
         |\Psi\rangle \in \mathcal{H}_P
      \right)
   \big]\mkern-4.3mu\big]}_v
   =
   1
   \;\;\;\;   .
\end{equation}
\smallskip

\noindent By contrast, suppose that $\{ \mathcal{H}_Q, \mathcal{H}_P\} \subseteq \mathcal{L}$ but $\mathcal{H}_Q$ and $\mathcal{H}_P$ are \textit{incomparable} with each other, i.e., $\mathcal{H}_Q \nleq \mathcal{H}_P$ and $\mathcal{H}_P \nleq \mathcal{H}_Q$. This means that $\mathcal{H}_Q \cap \mathcal{H}_P = \{0\}$, where $\cap$ denotes the set-theoretic operation of intersection, while $\{0\}$ is the zero-dimensional subspace, a proper subset of any closed linear subspace of $\mathcal{H}$. In such a case, the state $|\Psi\rangle \in \mathcal{H}_Q$ can be a member of $\mathcal{H}_P$ only if $|\Psi\rangle = 0$, where 0 is the zero-vector, the solitary vector contained in $\{0\}$. But, since any state $|\Psi\rangle$ of the physical system must differ from 0, it follows that the counterfactual truth-value of $P$ in $|\Psi\rangle \in \mathcal{H}_Q$ can be treated the same as the factual truth-value of $P$ in $|\Psi\rangle \notin \mathcal{H}_P$, that is,\smallskip

\begin{equation} \label{FALSE} 
   \{
      \mathcal{H}_Q
      ,
      \mathcal{H}_P
   \}
   \subseteq
   \mathcal{L}
   ,
   \>
   \mathcal{H}_Q
   \nleq
   \mathcal{H}_P
   ,
   \mathcal{H}_P
   \nleq
   \mathcal{H}_Q
   \textnormal{:}
   \;\;
   {\big[\mkern-4.3mu\big[ 
      P \left(
         |\Psi\rangle \in \mathcal{H}_Q
      \right)
   \big]\mkern-4.3mu\big]}_v
    =
   {\big[\mkern-4.3mu\big[ 
      P \left(
         |\Psi\rangle \notin \mathcal{H}_P
      \right)
   \big]\mkern-4.3mu\big]}_v
   =
   0
   \;\;\;\;   .
\end{equation}
\smallskip

\noindent To determine how a closed linear subspace $\mathcal{H}_P$ of a Hilbert space $\mathcal{H}$ can uniquely represent a proposition $P$, recall that $\mathcal{H}_P$ is \textit{the range} of the projection operator $\hat{P}$ (i.e., self-adjoint idempotent operator) acting on $\mathcal{H}$ \cite{Kalmbach}. Explicitly, $\mathcal{H}_P$ is the subset of the vectors $|\Psi\rangle \in \mathcal{H}$ that are in the image of the projection operator $\hat{P}$:\smallskip

\begin{equation}  
   \mathcal{H}_P
   \equiv
   \mathrm{ran}(\hat{P})
   =
   \left\{
      |\Psi\rangle \in \mathcal{H}
      \textnormal{:}
      \mkern10mu
      \hat{P} |\Psi\rangle
      =
      |\Psi\rangle
   \right\}
   \;\;\;\;   .
\end{equation}
\smallskip

\noindent As the set of the eigenvalues of $\hat{P}$ is contained in $\{0,1\}$, one can consider bijective correspondence between projection operators and propositions. Providing such correspondence, one gets an isomorphism from the set of propositions $\{P\}$ to the set of the ranges of the projection operators $\{\mathrm{ran}(\hat{P})\}$, explicitly,\smallskip

\begin{equation} \label{ISOM} 
   P
   \iff
   \mathrm{ran}(\hat{P})
   \;\;\;\;   .
\end{equation}
\smallskip

\noindent Recall that a set of two or more nontrivial projection operators on $\mathcal{H}$ (i.e., ones that differ from the identity operator $\hat{1}$ and the zero operator $\hat{0} = \hat{1} - \hat{1}$) is called \textit{a context} $\Sigma$\smallskip

\begin{equation}  
   \Sigma
   =
   \{\hat{P}\}
   \;\;\;\;   
\end{equation}
\smallskip

\noindent if any two members of $\Sigma$, say $\hat{P}_A$ and $\hat{P}_B$, are orthogonal to each other, i.e.,\smallskip

\begin{equation}  
   \hat{P}_A
   \hat{P}_B
   =
   \hat{P}_B
   \hat{P}_A
   =
   \hat{0}
   \;\;\;\;   ,
\end{equation}
\smallskip

\noindent and the resolution of identity is associated with $\Sigma$:\smallskip

\begin{equation}  
   \sum_{\hat{P} \in \Sigma}
   \hat{P}
   =
   \hat{1}
   \;\;\;\;   .
\end{equation}
\smallskip

\noindent In view of (\ref{ISOM}), one may also regard the context as the set of the compatible propositions $\Sigma = \{P\}$.\\

\noindent For a Hermitian operator with the continuous spectrum (such as the position $x$ and the momentum $p$), the projection operator $\hat{P}_U$, which relates to an interval $U = [u,u+\Delta u]$, takes the form\smallskip

\begin{equation}  
   \hat{P}_U
   \equiv
   \int_u^{u+\Delta u}
      \mathrm{d}u
      \,
      |u\rangle
      \langle u|
   \;\;\;\;   ,
\end{equation}
\smallskip

\noindent so that the projection operator $\hat{P}_{U^\prime}$ relating to another interval $U^\prime = [u^\prime,u^\prime+\Delta u^\prime]$ is orthogonal to $\hat{P}_U$ if and only if $U$ and $U^\prime$ do not intersect. Correspondingly, in that case the resolution of identity takes the form\smallskip

\begin{equation}  
   \int
      \mathrm{d}u
      \,
      |u\rangle
      \langle u|
      =
      1
   \;\;\;\;   .
\end{equation}
\smallskip

\noindent Since a subspace $\mathcal{P} \subseteq \mathcal{H}$ is \textit{invariant} under the projection operator $\hat{P}$ on $\mathcal{H}$ if the image of any vector $|\Psi\rangle$ in $\mathcal{P}$ under $\hat{P}$ remains within $\mathcal{P}$ and so\smallskip

\begin{equation}  
   \hat{P} \mathcal{P}
   =
   \left\{
      |\Psi\rangle \in \mathcal{P}
      \textnormal{:}
      \mkern10mu
      \hat{P} |\Psi\rangle
   \right\}
   \subseteq
   \mathcal{P}
   \;\;\;\;   ,
\end{equation}
\smallskip

\noindent the set of all invariant subspaces $\mathcal{P}$ of $\mathcal{H}$ invariant under the projection operator $\hat{P}$ is determined by\smallskip

\begin{equation}  
   \mathcal{L}(\hat{P})
   =
   \left\{
      \mathcal{P} \subseteq \mathcal{H}
      \textnormal{:}
      \mkern10mu
      \hat{P} \mathcal{P}
      \subseteq
      \mathcal{P}
   \right\}
   \;\;\;\;   .
\end{equation}
\smallskip

\noindent Consider the set of the invariant subspaces $\mathcal{L}(\Sigma)$ invariant under \textit{every} projection operator from the context $\Sigma$:\smallskip

\begin{equation}  
   \mathcal{L}(\Sigma)
   =
   \bigcap_{\hat{P} \in \Sigma}
   \mathcal{L}(\hat{P})
   \;\;\;\;   .
\end{equation}
\smallskip

\noindent The elements of this set form a complete lattice called \textit{the invariant-subspace lattice of the context} $\Sigma$ \cite{Radjavi}. It is straightforward to verify that each invariant-subspace lattice $\mathcal{L}(\Sigma)$ contains only mutually commuting subspaces (corresponding to mutually commutable projection operators), which means that each $\mathcal{L}(\Sigma)$ is a Boolean algebra.\\

\noindent Because $\hat{P}\{0\} \subseteq \{0\}$ and $\hat{P}\mathcal{H} \subseteq \mathcal{H}$ for all $\hat{P} \in \Sigma$, the subspaces $\{0\}$ and $\mathcal{H}$ are elements of each invariant-subspace lattice $\mathcal{L}(\Sigma)$.\\

\noindent Let $\mathcal{O} = \{\Sigma\}$ be the set of all the contexts associated with the quantum system. Then, the collection of the invariant-subspace lattices that is in one-to-one correspondence with  $\mathcal{O}$ can be defined as\smallskip

\begin{equation}  
   \left\{
      \mathcal{L}(\Sigma)
   \right\}
   \equiv
   \Big\{
      \Sigma \in \mathcal{O}
      :
      \;
      \mathcal{L}(\Sigma)
   \Big\}
   \;\;\;\;   .
\end{equation}
\smallskip

\noindent If all the lattices $\mathcal{L}(\Sigma)$ from the collection $\{\mathcal{L}(\Sigma)\}$ are \textit{pasted} (or stitched) together at their common elements (i.e., aside from identical elements, the subspaces $\{0\}$ and $\mathcal{H}$), namely,\smallskip

\begin{equation} \label{HL} 
   \mathcal{L}(\mathcal{H})
   =
   \bigcup_{\Sigma \in \mathcal{O}}
      \mathcal{L}({\Sigma})
   \;\;\;\;   ,
\end{equation}
\smallskip

\noindent where $\cup$ denotes the set-theoretic union carried out simultaneously on elements of $\{\mathcal{L}(\Sigma)\}$, then the resulted logic $\mathcal{L}(\mathcal{H})$ will be the Hilbert lattice, the complete orthomodular lattice based on all closed linear subspaces of the Hilbert space $\mathcal{H}$ \cite{Svozil08}. Thus, \textit{the Hilbert lattice $\mathcal{L}(\mathcal{H})$ is the union of the collection $\{\mathcal{L}(\Sigma)\}$}.\\

\noindent Within this union, any two subspaces $\mathcal{H}_Q$ and $\mathcal{H}_P$ -- regardless of whether they are elements of one invariant-subspace lattice or belong to two different lattices -- constitute a two-element subset of $\mathcal{L}(\mathcal{H})$; in symbols, $\{\mathcal{H}_Q, \mathcal{H}_P\} \subseteq \mathcal{L}(\mathcal{H})$. Consequently, within the structure of $\mathcal{L}(\mathcal{H})$, any two subspaces $\mathcal{H}_Q$ and $\mathcal{H}_P$ are either comparable or incomparable with each other, which means that the counterfactual truth-value of the proposition $P$ in the state $|\Psi\rangle \in \mathcal{H}_Q$ is always definite. In symbols,\smallskip

\begin{equation}  
   \forall
   \{
      \mathcal{H}_Q
      ,
      \mathcal{H}_P
   \}
   \subseteq
   \mathcal{L}(\mathcal{H})
   \textnormal{:}
   \mkern10mu
   \;\;
   {\big[\mkern-4.3mu\big[ 
      P \left(
         |\Psi\rangle \!\in\! \mathcal{H}_Q
      \right)
   \big]\mkern-4.3mu\big]}_v
   \in
   \{1,0\}
   \;\;\;\;   .
\end{equation}
\smallskip

\section{EPR paradox in the double-slit experiment}  

\noindent By reason of interaction, in Einstein’s version of the double-slit experiment, plate 1 and a particle become a combined system, which is associated with the tensor product\smallskip

\begin{equation}  
   \mathcal{H}^{\text{combined}}
   \equiv
   \mathcal{H}^{\text{plate1}}
   \otimes   
   \mathcal{H}^{\text{particle}}
   \;\;\;\;   
\end{equation}
\smallskip

\noindent of the two Hilbert spaces $\mathcal{H}^{\text{plate1}}$ and $\mathcal{H}^{\text{particle}}$ for plate 1 and the particle, respectively. The pure state of this combined system is described by\smallskip

\begin{equation}  
   |\Psi_{\text{combined}}\rangle
   \equiv
   |\Psi_{\text{plate1}}\rangle
   \otimes   
   |\Psi_{\text{particle}}\rangle
   \;\;\;\;   ,
\end{equation}
\smallskip

\noindent where $|\Psi_{\text{plate1}}\rangle$ and $|\Psi_{\text{particle}}\rangle$ denote the pure states of plate 1 and the particle correspondingly. After the measurement is performed on plate 1 -- of either position $x$ or momentum $p$ – the state $|\Psi_{\text{combined}}\rangle$ resides in one of the following closed linear subspaces:\smallskip

\begin{equation}  
   \mathcal{H}^{\text{combined}}_{X}
   \equiv
   \mathcal{H}^{\text{plate1}}_{X}
   \otimes   
   \mathcal{H}^{\text{particle}}_{X}
   \;\;\;\;   ,
\end{equation}

\begin{equation}  
   \mathcal{H}^{\text{combined}}_{P\pm}
   \equiv
   \mathcal{H}^{\text{plate1}}_{P\pm}
   \otimes   
   \mathcal{H}^{\text{particle}}_{P\pm}
   \;\;\;\;   ,
\end{equation}
\smallskip

\noindent where $\mathcal{H}^{\text{plate1}}_{X}$ is the subspace representing the proposition ``Position $x$ of plate 1 is in interval $X$'' (this proposition is symbolized by $P^{\text{plate1}}_{X}$); similarly, $\mathcal{H}^{\text{particle}}_{X}$ is the subspace that represents the proposition ``Position $x$ of the particle is in interval $X$'' (symbolized by $P^{\text{particle}}_{X}$). In the same way, $\mathcal{H}^{\text{plate1}}_{P\pm}$ and $\mathcal{H}^{\text{particle}}_{P\pm}$ are the subspaces representing respectively the propositions ``Momentum $p$ of plate 1 is positive (negative)'' and ``Momentum $p$ of the particle is positive (negative)'', symbolized by $P^{\text{plate1}}_{P\pm}$ and $P^{\text{particle}}_{P\pm}$ in that order. Subsequently, the subspace $\mathcal{H}^{\text{combined}}_{X}$ represents the proposition $(P^{\text{plate1}}_{X} \sqcap P^{\text{particle}}_{X})$, while the subspaces $\mathcal{H}^{\text{combined}}_{P+}$ and $\mathcal{H}^{\text{combined}}_{P-}$ represent the propositions $(P^{\text{plate1}}_{P+} \sqcap P^{\text{particle}}_{P+})$ and $(P^{\text{plate1}}_{P-} \sqcap P^{\text{particle}}_{P-})$.\\

\noindent Assume that immediately after the particle passes plate 1, one measures the position of this plate and finds $x \in X$. Upon doing so, one verifies the proposition $P^{\text{plate1}}_{X}$, which can be written down as\smallskip

\begin{equation}  
   {\Big[\mkern-4.9mu\Big[ 
      P^{\text{plate1}}_{X}
      \left(
         |\Psi_{\text{plate1}}\rangle
         \!\in\!
         \mathcal{H}^{\text{plate1}}_{X}
      \right)
   \Big]\mkern-4.9mu\Big]}_v
   =
   1
   \;\;\;\;   .
\end{equation}
\smallskip

\noindent Within the structure of the Hilbert lattice $\mathcal{L}(\mathcal{H}^{\text{plate1}})$, i.e., the union\smallskip

\begin{equation} \label{HP} 
   \mathcal{L}(\mathcal{H}^{\text{plate1}})
   =
   \bigcup_{\Sigma \in \mathcal{O}^{\text{plate1}}}
      \mathcal{L}({\Sigma})
   \;\;\;\;   ,
\end{equation}
\smallskip

\noindent where $\mathcal{O}^{\text{plate1}}$ denotes the set of all the contexts associated with plate 1, the subspace $\mathcal{H}^{\text{plate1}}_{X}$ is incomparable with both subspaces $\mathcal{H}^{\text{plate1}}_{P+}$ and $\mathcal{H}^{\text{plate1}}_{P-}$. In keeping with (\ref{FALSE}), this entails the following counterfactual truth-values of the propositions $P^{\text{plate1}}_{P+}$ and $P^{\text{plate1}}_{P-}$ in the state $|\Psi_{\text{plate1}}\rangle \in \mathcal{H}^{\text{plate1}}_{X}$:\smallskip

\begin{equation} \label{MX} 
   {\Big[\mkern-4.9mu\Big[ 
      P^{\text{plate1}}_{P\pm}
      \left(
         |\Psi_{\text{plate1}}\rangle
         \!\in\!
         \mathcal{H}^{\text{plate1}}_{X}
      \right)
   \Big]\mkern-4.9mu\Big]}_v
   =
   0
   \;\;\;\;   .
\end{equation}
\smallskip

\noindent After the measurement of the position of plate 1, the combined system is in the state $|\Psi_{\text{combined}}\rangle$ lying in $\mathcal{H}^{\text{combined}}_{X}$. Within the structure of the Hilbert lattice $\mathcal{L}(\mathcal{H}^{\text{combined}})$, namely,\smallskip

\begin{equation} \label{HC} 
   \mathcal{L}(\mathcal{H}^{\text{combined}})
   =
   \bigcup_{\Sigma \in \mathcal{O}^{\text{combined}}}
      \mathcal{L}({\Sigma})
   \;\;\;\;   ,
\end{equation}
\smallskip

\noindent where $\mathcal{O}^{\text{combined}}$ denotes the set of all the contexts associated with the combined system, the subspace $\mathcal{H}^{\text{combined}}_{X}$ is incomparable with both subspaces $\mathcal{H}^{\text{combined}}_{P+}$ and $\mathcal{H}^{\text{combined}}_{P-}$. Accordingly, in the product state $|\Psi_{\text{combined}}\rangle \in \mathcal{H}^{\text{combined}}_{X}$ one must assign the counterfactual value of false to the propositions $(P^{\text{plate1}}_{P\pm} \sqcap P^{\text{particle}}_{P\pm})$:\smallskip

\begin{equation} \label{MMX} 
   {\Big[\mkern-4.9mu\Big[ 
      \left(
      P^{\text{plate1}}_{P\pm} \sqcap P^{\text{particle}}_{P\pm}
      \right)
      \left(
         |\Psi_{\text{combined}}\rangle
         \!\in\!
         \mathcal{H}^{\text{combined}}_{X}
      \right)
   \Big]\mkern-4.9mu\Big]}_v
   =
   0
   \;\;\;\;   .
\end{equation}
\smallskip

\noindent Then again, in the product state $|\Psi_{\text{combined}}\rangle \in \mathcal{H}^{\text{combined}}_{X}$ those propositions are equivalent to the disjunctions of the propositions and so\smallskip

\begin{equation}  
   {\Big[\mkern-4.9mu\Big[ 
      P^{\text{plate1}}_{P\pm}
      \left(
         |\Psi_{\text{plate1}}\rangle
         \!\in\!
         \mathcal{H}^{\text{plate1}}_{X}
      \right)
      \sqcap
      P^{\text{particle}}_{P\pm}
      \left(
         |\Psi_{\text{particle}}\rangle
         \!\in\!
         \mathcal{H}^{\text{particle}}_{X}
      \right)
   \Big]\mkern-4.9mu\Big]}_v
   =
   0
   \;\;\;\;   .
\end{equation}
\smallskip

\noindent If the truth values 0 and 1 are interpreted as integers, this can be expressed with the ordinary operation of multiplication $\times$; as a result, the statement (\ref{MMX}) can be rewritten as\smallskip

\begin{equation}  
   {\Big[\mkern-4.9mu\Big[ 
      P^{\text{plate1}}_{P\pm}
      \left(
         |\Psi_{\text{plate1}}\rangle
         \!\in\!
         \mathcal{H}^{\text{plate1}}_{X}
      \right)
   \Big]\mkern-4.9mu\Big]}_v
   \times
   {\Big[\mkern-4.9mu\Big[ 
      P^{\text{particle}}_{P\pm}
      \left(
         |\Psi_{\text{particle}}\rangle
         \!\in\!
         \mathcal{H}^{\text{particle}}_{X}
      \right)
   \Big]\mkern-4.9mu\Big]}_v
   =
   0
   \;\;\;\;   .
\end{equation}
\smallskip

\noindent Due to (\ref{MX}), it is consistent with\smallskip

\begin{equation}  
   {\Big[\mkern-4.9mu\Big[ 
      P^{\text{particle}}_{P+}
      \left(
         |\Psi_{\text{particle}}\rangle
         \!\in\!
         \mathcal{H}^{\text{particle}}_{X}
      \right)
   \Big]\mkern-4.9mu\Big]}_v
   \times
   {\Big[\mkern-4.9mu\Big[ 
      P^{\text{particle}}_{P-}
      \left(
         |\Psi_{\text{particle}}\rangle
         \!\in\!
         \mathcal{H}^{\text{particle}}_{X}
      \right)
   \Big]\mkern-4.9mu\Big]}_v
   =
   0
   \;\;\;\;    
\end{equation}
\smallskip

\noindent and may entail the equivalence:\smallskip

\begin{equation} \label{EQ} 
   {\Big[\mkern-4.9mu\Big[ 
      P^{\text{particle}}_{X}
      \left(
         |\Psi_{\text{particle}}\rangle
         \!\in\!
         \mathcal{H}^{\text{particle}}_{X}
      \right)
   \Big]\mkern-4.9mu\Big]}_v
   =
   {\Big[\mkern-4.9mu\Big[ 
      P^{\text{particle}}_{P\pm}
      \left(
         |\Psi_{\text{particle}}\rangle
         \!\in\!
         \mathcal{H}^{\text{particle}}_{X}
      \right)
   \Big]\mkern-4.9mu\Big]}_v
   =
   1
   \;\;\;\;   .
\end{equation}
\smallskip

\noindent Let’s analyze the meaning of this equivalence. The truth of the proposition $P^{\text{particle}}_{X}$ signifies the interference pattern. Indeed, as it is shown in the paper \cite{Zurek}, all particles that start at the position $x \in X$ form a sub-ensemble whose contribution to the total interference pattern on plate 3 is perfect. On the other hand, the truth of the proposition $P^{\text{particle}}_{P+}$ or $P^{\text{particle}}_{P-}$ indicates a fairly good knowledge of the path of each particle: If $P^{\text{particle}}_{P+}$ is true, then one will guess that the particle passed through slit A; and if $P^{\text{particle}}_{P-}$ is true, then the particle apparently passed through slit B. Thus, the equivalence (\ref{EQ}) means that \textit{it is possible to know the path of each particle in the sub-ensemble $|\Psi_{\textnormal{particle}}\rangle \in \mathcal{H}^{\textnormal{particle}}_{X}$ without disturbing the interference pattern}. This contradicts quantum mechanics which predicts that the measurement of the position of a particle destroys any information about its momentum, and vice versa.\\

\noindent It is not difficult to notice that the paradoxical equivalence (\ref{EQ}) is similar to the EPR paradox \cite{EPR}; in view of that, the authors of \cite{Zurek} termed Einstein's design ``the EPR paradox in the double-slit experiment''.\\

\section{How to resolve the paradox}  

\noindent Consider ways to resolve the EPR paradox in the double-slit experiment.\\

\noindent One may assume that the following correspondence takes place:\smallskip

\begin{equation} \label{CORR} 
   {\Big[\mkern-4.9mu\Big[ 
      P^{\text{plate1}}_{P\pm}
      \left(
         |\Psi_{\text{plate1}}\rangle
         \!\in\!
         \mathcal{H}^{\text{plate1}}_{X}
      \right)
   \Big]\mkern-4.9mu\Big]}_v
   =
   {\Big[\mkern-4.9mu\Big[ 
      P^{\text{particle}}_{P\pm}
      \left(
         |\Psi_{\text{particle}}\rangle
         \!\in\!
         \mathcal{H}^{\text{particle}}_{X}
      \right)
   \Big]\mkern-4.9mu\Big]}_v
   \;\;\;\;   .
\end{equation}
\smallskip

\noindent According to it, the propositions $P^{\text{particle}}_{P+}$ and $P^{\text{particle}}_{P-}$ must have the value of false in the state $|\Psi_{\text{particle}}\rangle \in \mathcal{H}^{\text{particle}}_{X}$ because the propositions $P^{\text{plate1}}_{P+}$ and $P^{\text{plate1}}_{P-}$ take on the value of false in the state $|\Psi_{\text{plate1}}\rangle \in \mathcal{H}^{\text{plate1}}_{X}$. This puts an end to the paradoxical equivalence (\ref{EQ}):\smallskip

\begin{equation}  
   {\Big[\mkern-4.9mu\Big[ 
      P^{\text{particle}}_{X}
      \left(
         |\Psi_{\text{particle}}\rangle
         \!\in\!
         \mathcal{H}^{\text{particle}}_{X}
      \right)
   \Big]\mkern-4.9mu\Big]}_v
   \neq
   {\Big[\mkern-4.9mu\Big[ 
      P^{\text{particle}}_{P\pm}
      \left(
         |\Psi_{\text{particle}}\rangle
         \!\in\!
         \mathcal{H}^{\text{particle}}_{X}
      \right)
   \Big]\mkern-4.9mu\Big]}_v
   \;\;\;\;   .
\end{equation}
\smallskip

\noindent Be it as it may, in view of the fact that the measurements (of the position or the momentum) on plate 1 are always performed after the particle has interacted with it, the correspondence (\ref{CORR}) implies that the particle ``knows'' that one decided to measure the position of plate 1 even when the particle is separated from it by a large distance. In other words, the correspondence (\ref{CORR}) implies that \textit{the principle of locality} is not applicable in quantum mechanics.\\

\noindent Alternatively, one may argue that the counterfactual truth-values (\ref{MX}) and (\ref{MMX}) are not admissible because they contradict the Kolmogorov axioms \cite{Kolmogorov}.\\

\noindent To be sure, consider the probabilities that the propositions $P^{\text{plate1}}_{P+}$ and $P^{\text{plate1}}_{P-}$ are verified in the state $|\Psi_{\text{plate1}}\rangle \in \mathcal{H}^{\text{plate1}}_{X}$: In keeping with (\ref{MX}), one finds:\smallskip

\begin{equation}  
   \Pr
   \Big[
      P^{\text{plate1}}_{P\pm}
      \left(
         |\Psi_{\text{plate1}}\rangle
         \!\in\!
         \mathcal{H}^{\text{plate1}}_{X}
      \right)
   \Big]
   =
   0
   \;\;\;\;  .
\end{equation}
\smallskip

\noindent But, if the sample space for momentum $p$ of plate 1 is $\Omega = \{\text{positive} ,\text{negative}\}$, Kolmogorov's axioms will imply\smallskip

\begin{equation}  
   \Pr
   \Big[
      P^{\text{plate1}}_{P+}
      \left(
         |\Psi_{\text{plate1}}\rangle
         \!\in\!
         \mathcal{H}^{\text{plate1}}_{X}
      \right)
   \Big]
   +
   \Pr
   \Big[
      P^{\text{plate1}}_{P-}
      \left(
         |\Psi_{\text{plate1}}\rangle
         \!\in\!
         \mathcal{H}^{\text{plate1}}_{X}
      \right)
   \Big]
   =
   1
   \;\;\;\;  .
\end{equation}
\smallskip

\noindent Thus, (\ref{MX}), as well as (\ref{MMX}), must be nonadmissible, and, as a result, instead of (\ref{EQ}) one can find only\smallskip

\begin{equation}  
   {\Big[\mkern-4.9mu\Big[ 
      P^{\text{particle}}_{X}
      \left(
         |\Psi_{\text{particle}}\rangle
         \!\in\!
         \mathcal{H}^{\text{particle}}_{X}
      \right)
   \Big]\mkern-4.9mu\Big]}_v
   =
   1
   \;\;\;\;   .
\end{equation}
\smallskip

\noindent However, the admissibility poses an additional problem: Why should the Kolmogorov axioms be relevant for truth-values of propositions? Really, unlike propositions, probability assignments are neither true nor false; that is, probability assignments are not propositions. So, any justification of Kolmogorov’s axiomatic system for truth-values will require further assumptions that bring probabilities into a logic. Putting it differently, without appending more hypotheses, one cannot rationalize the requirement of admissibility for truth-values of propositions.\\

\section{Giving up the condition of the Hilbert lattice}  

\noindent Unlike the Hilbert lattice $\mathcal{L}(\mathcal{H})$, the collection of the invariant-subspace lattices $\{\mathcal{L}(\Sigma)\}$ is a structure where a pair of subspaces $\mathcal{H}_Q$ and $\mathcal{H}_P$ can be \textit{neither comparable nor incomparable} with each other.\\

\noindent To be sure, suppose that subspaces $\mathcal{H}_Q$ and $\mathcal{H}_P$ are elements belonging respectively to the posets $\mathcal{L}(\Sigma_Q)$ and $\mathcal{L}(\Sigma_P)$ whose intersection $\mathcal{L}(\Sigma_Q) \cap \mathcal{L}(\Sigma_P)$ contains neither $\mathcal{H}_Q$ nor $\mathcal{H}_P$. In that case, within the structure of the collection $\{\mathcal{L}(\Sigma)\} = \{\mathcal{L}(\Sigma_Q), \mathcal{L}(\Sigma_P), \dots\}$, there is no poset that contains both $\mathcal{H}_Q$ and $\mathcal{H}_P$, which means that \textit{an ordering} -- i.e., a binary relation over a pair of elements from one poset -- has no meaning for $\mathcal{H}_Q$ and $\mathcal{H}_P$ at all.\\

\noindent Hence, giving up the condition of the Hilbert lattice, i.e., the union of the collection $\{\mathcal{L}(\Sigma)\}$, the counterfactual truth-value of the proposition $P$ in the state $|\Psi\rangle \in \mathcal{H}_Q$ becomes meaningless, i.e., without the truth values. Symbolically, this can be presented by\smallskip

\begin{equation}  
   \{
      \mathcal{H}_Q
      ,
      \mathcal{H}_P
   \}
   \nsubseteq
   \mathcal{L}(\Sigma)
   \in
   \left\{
      \mathcal{L}(\Sigma_Q), \mkern5mu \mathcal{L}(\Sigma_P), \mkern5mu \dots
   \right\}
   \textnormal{:}
   \mkern10mu
   \;\;
   {\big[\mkern-4.3mu\big[ 
      P \left(
         |\Psi\rangle \!\in\! \mathcal{H}_Q
      \right)
   \big]\mkern-4.3mu\big]}_v
   =
   \frac{0}{0}
   \;\;\;\;   ,
\end{equation}
\smallskip

\noindent where $\frac{0}{0}$ denotes an indeterminate value.\\

\noindent As follows, the relinquishment of the Hilbert lattice condition automatically renders the resolution of the EPR paradox in the double-slit experiment.\\

\noindent Indeed, within the structure of the collection of the invariant-subspace lattices associated with plate 1, namely,\smallskip

\begin{equation}  
   \left\{
      \mathcal{L}(\Sigma^{\text{plate1}})
   \right\}
   \equiv
   \left\{
      \mathcal{L}\left(\Sigma^{\text{plate1}}_X\right)
      ,
      \mkern5mu
      \mathcal{L}\left(\Sigma^{\text{plate1}}_P\right)
      ,\mkern5mu
      \dots
   \right\}
   \;\;\;\;  ,
\end{equation}
\smallskip

\noindent the subspace $\mathcal{H}^{\text{plate1}}_X$ from the poset $\mathcal{L}(\Sigma^{\text{plate1}}_X)$ is neither comparable nor incomparable with the subspaces $\mathcal{H}^{\text{plate1}}_{P+}$ and $\mathcal{H}^{\text{plate1}}_{P-}$ from the poset $\mathcal{L}(\Sigma^{\text{plate1}}_P)$. This means that within this structure, the counterfactual truth-values of the propositions $P^{\text{plate1}}_{P+}$ and $P^{\text{plate1}}_{P-}$ in the state $|\Psi_{\text{plate1}}\rangle \in \mathcal{H}^{\text{plate1}}_X$ have no meaning:\smallskip

\begin{equation}  
   {\Big[\mkern-4.9mu\Big[ 
      P^{\text{plate1}}_{P\pm} 
      \left(
         |\Psi_{\text{plate1}}\rangle \!\in\! \mathcal{H}^{\text{plate1}}_X
      \right)
   \Big]\mkern-4.9mu\Big]}_v
   =
   \frac{0}{0}
   \;\;\;\;   .
\end{equation}
\smallskip

\noindent In the same way, within the structure of the collection of the invariant-subspace lattices associated with the combined system, i.e.,\smallskip

\begin{equation}  
   \left\{
      \mathcal{L}(\Sigma^{\text{combined}})
   \right\}
   \equiv
   \left\{
      \mathcal{L}\left(\Sigma^{\text{combined}}_X\right)
      ,
      \mkern5mu
      \mathcal{L}\left(\Sigma^{\text{combined}}_P\right)
      ,\mkern5mu
      \dots
   \right\}
   \;\;\;\;  ,
\end{equation}
\smallskip

\noindent the subspace $\mathcal{H}^{\text{combined}}_X \in \mathcal{L}(\Sigma^{\text{combined}}_X)$ is neither comparable nor incomparable with the subspaces $\mathcal{H}^{\text{plate1}}_{P\pm} \otimes \mathcal{H}^{\text{particle}}_{P\pm} \in \mathcal{L}(\Sigma^{\text{combined}}_P)$; therefore,\smallskip

\begin{equation}  
   {\Big[\mkern-4.9mu\Big[ 
      \left(
         P^{\text{plate1}}_{P\pm}
         \sqcap
         P^{\text{particle}}_{P\pm}
      \right)
      \left(
         |\Psi_{\text{combined}}\rangle \!\in\! \mathcal{H}^{\text{combined}}_X
      \right)
   \Big]\mkern-4.9mu\Big]}_v
   =
   \frac{0}{0}
   \;\;\;\;   .
\end{equation}
\smallskip

\noindent This gives\smallskip

\begin{equation}  
   {\Big[\mkern-4.9mu\Big[ 
      P^{\text{particle}}_{P\pm}
      \left(
         |\Psi_{\text{particle}}\rangle \!\in\! \mathcal{H}^{\text{particle}}_X
      \right)
   \Big]\mkern-4.9mu\Big]}_v
   =
   \frac{0}{0}
   \;\;\;\;   .
\end{equation}
\smallskip

\noindent Accordingly, with no condition of the Hilbert lattice, i.e., without (\ref{HP}) and (\ref{HC}), the paradox (\ref{EQ}) does not emerge.\\

\section{Conclusion remarks}  

\noindent In Copenhagen interpretation of quantum mechanics, the nonequivalence of the truth-values of the propositions $P_{\text{wave-like}}$  and $P_{\text{particle-like}}$ is simply postulated. Such postulate is of great importance: Complementarity is the chief foundation on which the orthodox account of quantum theory is built. According to Bohr, the demand of complementarity in quantum mechanics is logically on a par with the requirement of relativity in theory of relativity \cite{Faye}.\\

\noindent The question that one may ask is this: Can Bohr's complementarity principle be emergent? Specifically, can the nonequivalence (\ref{CP}) be derived?\\

\noindent As it follows from the empiric logic analysis of Einstein's modification of the double-slit experiment, within the structure of the Hilbert lattice, the nonequivalence of the truth-values of $P_{\text{wave-like}}$  and $P_{\text{particle-like}}$ might be explained by the violation of the principle of locality, or by the introduction of the probabilistic axioms into a logic. However, both such possibilities have a clear physical content which means that they merely put forward new physical hypotheses in place of the complementarity principle.\\

\noindent In contrast to this, when the condition of the Hilbert lattice is dropped, the nonequivalence (\ref{CP}) arises from the Hilbert space formalism without additional hypotheses. One can conclude from here that the Bohr's complementarity principle is a result of truth-value gaps which are intrinsic to the structure of the collection of the invariant-subspace lattices $\{\mathcal{L}(\Sigma)\}$. In other words, this principle is a consequence of the fact that within $\{\mathcal{L}(\Sigma)\}$ the valuation relations from propositions to the set $\{1,0\}$ are not total functions.\\

\bibliographystyle{References}
\bibliography{D_slit}

\end{document}